\documentclass[12pt,fleqn]{article}
\usepackage{amsmath}
\usepackage{amsfonts}
\usepackage{epsfig}
\newlength{\dinwidth}
\newlength{\dinmargin}
\newcommand{\f}[2]{\frac{#1}{#2}}

\def\ra{\rightarrow}

\def\ep{\varepsilon}

\def\k{{\bf k}}
\def\as{\alpha_s}
\def\asb{\bar{\alpha}_s}
\def\q{{\bf q}}

\def\bks{\!\!\!\!\!\!\!\!\!}

\def\o{\omega}
\def\g{\gamma}
\def\G{\Gamma}
\def\op{\o_{\mathbb{P}}} 
\newcommand{\be}{\begin{equation}}
\newcommand{\ee}{\end{equation}}
\newcommand{\bea}{\begin{align}}
\newcommand{\eea}{\end{align}}
\newcommand{\nn}{\nonumber}

\begin{document}

\title{Irreducible part of the next-to-leading BFKL kernel}
\author{G. Camici and M. Ciafaloni\footnotemark{*}\\
{\em Dipartimento di Fisica, Universit\`a di Firenze} \\
{\em and INFN, Sezione di Firenze} \\
{\em Largo E. Fermi, 2 - 50125  Firenze}}
\date{}
\maketitle
\thispagestyle{empty}
\begin{abstract}
On the basis of previous work by Fadin, Lipatov, and collaborators, and of
our group, we extract the "irreducible" part of the next-to-leading (NL) BFKL
kernel, we compute its (IR finite) eigenvalue function, and we discuss its
implications for small-$x$ structure functions.
We find consistent running coupling effects and sizeable NL corrections
to the Pomeron intercept and to the gluon anomalous dimension.
The qualitative effect of such corrections
is to smooth out the small-$x$ rise of structure functions
at low values of $Q^2$. A more quantitative analysis 
will be possible after the extraction of some additional, energy-scale 
dependent contributions to the kernel, which are not treated here.
\end{abstract}

\begin{center}
\end{center}
\vspace*{2.5 cm}
{\small $~^*$ email: camici@fi.infn.it, ciafaloni@fi.infn.it}
\newpage
\setcounter{page}{1}

The small-$x$ rise of structure functions at HERA \cite{1} has stimulated 
an impressive theoretical effort [2-15] in order to understand the
high-energy behaviour in QCD.

In a series of papers [2-3,5-9,11], Fadin, Lipatov and collaborators have 
investigated the high-energy cluster expansion (Fig. 1) of the parton-parton 
cross-section, with the purpose of generalizing the leading-$\log s$ BFKL
equation\cite{2} to the next-to-leading (NL) order.
Similar results for the (heavy) $q\bar{q}$ production cross section\cite{4}
and eigenvalues \cite{4,12}, and for the squared gluon emission amplitude
\cite{10} have been produced by other authors.

The outcome of such an analysis is the calculation of some "irreducible" vertices
which, defined by gluon Regge-pole factorization \cite{2}, have the role of 
incorporating the low energy features of the QCD scattering amplitudes.
They are, more precisely, the two-loop gluon trajectory renormalization
\cite{8}, the one-loop reggeon-reggeon-gluon ($RRG$) vertex [5-7] and the 
$RRQ\bar{Q}$ \cite{4} and $RRGG$\cite{9,11} clusters at tree-level
(Fig. 1 (d-f)).

Such vertices, which involve parton counting in the final state, suffer 
from mass singularities and need be combined in a sum, with parton number 
$n=0,1,2~$, in order to define the IR finite, irreducible part of the
NL kernel.
This cancellation of singularities was shown in Ref. \cite{15}.

Recently, the $N_f$-dependent (or $q\bar{q}$) part of the kernel was 
extracted by the authors 
\cite{12,13},its eigenvalue was computed, and the ensuing 
Pomeron shift and anomalous dimensions were evaluated. Both shift and 
resummation effects turn out to be suppressed, in this case, by a 
nonplanar colour factor, which needs not be present in 
the gluonic case.

Interest in extracting the full gluonic contribution is thus substantial, 
but is not as simple as in the $q\bar{q}$ case, however. In fact, in order
to define the irreducible vertices in the cluster expansion, we need to
subtract the leading kernel iteration with $\log s$ accuracy.
In addition, we need to define an off-shell scale for the energy, starting 
from a partonic cross-section which is not an IR safe hard process and thus
suffers from Coulomb-like and possibly collinear singularities due to
initial partons.

In performing this procedure, we shall distinguish two kinds of contributions
to the kernel: (i) the properly irreducible ones, depicted in Fig. 1, 
which come from the $RR$, $RRG$ and $RRGG$ clusters to be defined below, and 
(ii) some additional contributions, that we call
{\em energy-scale dependent},
which are remainders of the leading term, with its energy scale, after
the required factorization of Coulomb and collinear singularities.
The last step involves the choice of a
factorization scheme of vertices and kernel
which should allow the use of the latter in hard processes.

The purpose of the present note is to perform in part
this program, by combining the "irreducible" terms mentioned above
and by discussing their eigenvalues and related
features. This will allow us to understand
the running coupling effects and the
main consequences for high energies and anomalous dimension behaviour.
On the other hand, full quantitative results can
be obtained only after the extraction
of the energy scale dependent terms,
which is deferred to a subsequent analysis.

Let us start by defining the irreducible terms more precisely.
We work in $\o$-space
(Mellin transform in the energy variable $s$) and
transverse momentum space with
respect to the incoming partons' axis.
The leading kernel in $D=4+2\ep$ dimensions
has the well-known \cite{2} form
\begin{equation}
\bks K^{(L)}=\f{\asb}{\o}K_0(\k_1,\k_2)=\f{\asb}{\o\G(1-\ep)}\f{1}{(\k_1-\k_2)^2}+
\f{2\o^{(1)}(\k_1^2)}{\o}\pi^{1+\ep}(\mu^2)^\ep\delta^{2(1+\ep)}(\k_1-\k_2),
\end{equation}
where we have adopted the notation of Fig. 1  (i. e., $\k$'s ($\q$'s) for the 
exchanged (emitted) momenta),
and we have introduced the one-loop gluon trajectory
\cite{2,8} 
\begin{equation}
\o^{(1)}(\k^2)=-\f{\asb}{4}C(\ep)\left(\f{\k^2}{\mu^2}\right)^\ep,~~~~~~~~~~~
\Big( C(\ep)\equiv \f{2}{\ep}\f{\G^2(1+\ep)}{\G(1+2\ep)}\Big),
\end{equation}
and the notation
\begin{equation}
\asb=\f{N_c\as}{\pi},~~~~~~\as(\mu^2)=\f{g_\mu^2}{(4\pi)^{(1+\ep)}}\G(1-\ep).
\end{equation}
We also understand that transverse integrations carry the
measure $d[\k]\equiv d^{2(1+\ep)}\k/$ $\pi^{1+\ep}(\mu^2)^\ep$.

The kernel (1) is the prototype of the $\ep$-dependent kernels to
be written out.
It is finite for $\ep\ra 0$ and $\q\equiv \k_1-\k_2\neq 0$,
but may still have singular
eigenvalues because of the $\q=0$ singularity. The virtual term -
in this case the one-loop gluon trajectory -
regularizes the singularity by providing a subtraction
which yields a finite eigenvalue.

In fact by applying the kernel (4) to the test function $(\k^2_2)^{\g-1}$, 
with $0<\g<1$, and by using the integral
\begin{align}
\bks\bks\int\f{d[\k_2]}{\G(1-\ep)\q^2}\left(\f{\k_2^2}{\k_1^2}\right)^{\g-1}&=
\left(\f{\k_1^2}{\mu^2}\right)^\ep\f{1}{\ep}
\f{\G(1+\ep)\G(\g+\ep)\G(1-\g-\ep)}{\G(1-\ep)\G(\g+2\ep)\G(1-\g)}\nn\\
&=\left(\f{\k_1^2}{\mu^2}\right)^\ep\f{1}{\ep}
\left[\exp(\ep\chi_0(\g))+\f{1}{2}\ep^2(\psi^\prime(1-\g)-3\psi^\prime(\g))+
O(\ep^3)\right]
\end{align}
it is easy to combine real and virtual terms to obtain the
characteristic function
\begin{equation}
\chi_0(\g)=2\psi(1)-\psi(\g)-\psi(1-\g),
\end{equation}
where $\psi(\g)$ is the logarithmic derivative of the $\G$-function.

The NL kernel contains virtual and real emission terms also. The virtual term is 
the two-loop gluon trajectory
\begin{equation}
K_V^{(NL)}=\f{2\o^{(2)}(\k_1^2)}{\o}\pi(\pi\mu^2)^\ep\delta^{2(1+\ep)}(\q),
\end{equation}
which, by collecting the gluonic contributions only, is given by \cite{8}
\begin{align}
&2\o^{(2)}(\k_1^2)=\f{\asb^2}{4\ep^2}\left[
-\left(\f{\k_1^2}{\mu^2}\right)^\ep\f{11}{3}
\left(1-\f{\pi^2}{6}\ep^2\right)+\right.\nn\\
&\left.\left(\f{\k_1^2}{\mu^2}\right)^{2\ep}
\left(\f{11}{6}+\left(\f{\pi^2}{6}-\f{67}{18}\right)\ep+
\left(\f{202}{27}-\f{11\pi^2}{18}-\zeta(3)\right)\ep^2\right)\right].
\end{align}
The subtraction of the leading
term ($\sim2(\o^{(1)}\log s)^2$) is in this case
unambiguous, because the only available scale is $\k_1^2$.

The real emission terms involve several scales, and thus some prescription is 
required for the subtraction of the leading term.
We start by writing the one-gluon emission
amplitude with the one-loop corrections of
Ref. [5-7] as follows
\begin{align}
&M_{\ep a}^{(1)}=M_{\ep a}^{(0)}\left(1+\beta_A^{(1)}(\k_1^2)
+\beta_B^{(1)}(\k_2^2)+\right.\nn\\
&\left.\f{1}{2}\o^{(1)}(\k_1^2)\left(\log\f{s_1}{\q^2}+
\log\f{s_1}{\sqrt{\k_1^2\k_2^2}}\right)+\f{1}{2}\o^{(1)}(\k_2^2)
\left(\log\f{s_2}{\q^2}+\log\f{s_2}{\sqrt{\k_1^2\k_2^2}}\right)
\right)+\tilde{M}^{(1)}_{\ep a}
\end{align}
where $s_1,s_2$ are subenergy variables of the emitted gluon $q^\mu$
of polarization
$\ep$ and colour $a$, $\beta_A^{(1)}$($\beta_B^{(1)}$) are the one-loop 
corrections \cite{5,6} to the A (B) vertex, $M^{(0)}$ is the leading
amplitude, and $\tilde{M}^{(1)}$ is defined to be the NL irreducible one.
They are given by
\begin{equation}
\bks M^{(0)}_{\ep a}=\f{s}{\k_1^2\k_2^2}T_a~2g_\mu\ep\cdot J,~~~~~
J^\mu(q)=-k_1^\mu-k_2^\mu+\f{p_A^\mu}{p_A\cdot q}(\q^2-\k_1^2)-
\f{p_B^\mu}{p_B\cdot q}(\q^2-\k_2^2),
\end{equation}
and by
\begin{equation}
Re ~~\tilde{M}^{(1)}_{\ep a}=
\f{s}{\k_1^2\k_2^2}T_a\left[(2g_\mu\ep\cdot J)\f{\asb}{4}
\left[-\f{C(\ep)}{2}\f{\pi\cos\pi\ep}
{\sin\pi\ep}\left(\f{\q^2}{\mu^2}\right)^\ep+
\f{11}{6\ep}+\right.\right.\nn \end{equation}
\begin{equation}
\left.+\f{\q^2}{3}\f{\k_1^2+\k_2^2}{(\k_1^2-\k_2^2)^2}+
\left(\f{11}{6}\f{\k_1^2+\k_2^2}{\k_1^2-\k_2^2}-
\f{2}{3}\q^2\f{\k_1^2\k_2^2}{(\k_1^2-\k_2^2)^2}
\right)\log\f{\k_1^2}{\k_2^2}\right]+
\nn   \end{equation}
\begin{equation}
+(2g_\mu\ep\cdot J_s)\f{\asb}{4}\left[\f{\k_1^2\k_2^2}{3(\k_1^2-\k_2^2)}
\left(11+\q^2\f{2\q^2-\k_1^2-\k_2^2}{(\k_1^2-\k_2^2)^2}\right)
\log\f{\k_1^2}{\k_2^2}+
\right.\nn    \end{equation}
\begin{equation}
\left.\left.+\f{\q^2}{6}\left(1-(2\q^2-\k_1^2-\k_2^2)
\f{\k_1^2+\k_2^2}{(\k_1^2-\k_2^2)^2}\right)\right]\right],
\end{equation}
where $J^\mu$ is the gluon-emission current associated with high energy
scattering \cite{2},
while $J_s^\mu=\f{p_A^\mu}{p_A\cdot q}-\f{p_B^\mu}{p_B\cdot q}$
is the soft insertion current, occurring in $\tilde{M}^{(1)}$ only.

Notice that we have used in Eq. (10)
the small-$\q^2$ behaviour of Eq. (21) of Ref.
\cite{7}, together with the fixed $\q^2$ form of Eq. (86) of Ref. \cite{5}.
Furthermore, we have incorporated the $\o^{(1)}\log\k_i^2$
terms in the definition of
the {\em leading} part in Eq. (8), so as to subtract them out in a scale
invariant form.

The one-gluon contribution to the NL kernel comes from the interference of
$\tilde{M}^{(1)}$ in Eq. (10) with $M^{(0)}$ in Eq. (9).
Using the polarization sums
\begin{equation}
-J^2=\f{4\k_1^2\k_2^2}{\q^2},~~~~~~~~~
J\cdot J_s=\f{4\k_1\cdot\k_2}{\q^2},
\end{equation}
and performing, for simplicity, an azimuthal average of the polynomial
part in $\q^2$, we obtain
\begin{align}
K_{1g}^{(NL)}&=\f{\asb}{\o}\f{\asb}{4}\left[-\f{C(\ep)}{\G(1-\ep)}
\f{(\q^2/\mu^2)^\ep}{\q^2}\f{\pi\cos\pi\ep}{\sin\pi\ep}+\right.\nn\\
&\left. +\f{11}{3}\left(\f{1}{\ep\G(1-\ep)\q^2}+\f{1}{\k_1^2-\k_2^2}
\log\f{\k_1^2}{\k_2^2}\right)\right].
\end{align}

Finally, the two-gluon emission cluster has been recently computed in Ref
\cite{11}, where the authors suggest subtracting the leading term
\begin{equation}
\asb^2\int\f{d[\q_1]}{\q_1^2(\q-\q_1)^2}\int\f{dx}{x(1-x)}
\end{equation}
with a scale invariant rapidity phase space
$\int\f{dx}{x(1-x)}=2\log(1/\delta)$. By using this prescription,
we obtain, from Eq. (20) of Ref. \cite{11}, the expression
\begin{align}
K_{2g}^{(NL)}&=\f{\asb}{\o}\f{\asb}{4}\left[
\f{C(\ep)(\q^2/\mu^2)^\ep}{\G(1-\ep)\q^2}\left[
\f{1}{\ep}-\f{11}{6}+\left(\f{67}{18}-\f{\pi^2}{2}\right)\ep-
\left(\f{202}{27}-7\zeta(3)\right)\ep^2\right]\right.+\nn\\
&\left. -H_{coll}(\k_1,\k_2)+\tilde{H}(\k_1,\k_2)\right],
\end{align}
where we have introduced the "collinear" kernel
\begin{align}
&H_{coll}(\k_1,\k_2)=\f{1}{32}\left[2\left(\f{1}{\k_1^2}+\f{1}{\k_2^2}\right)+
\left(\f{1}{\k_2^2}-\f{1}{\k_1^2}\right)\log\f{\k_1^2}{\k_2^2}+
\left(118-\f{\k_1^2}{\k_2^2}-\f{\k_2^2}{\k_1^2}\right)\times\right.\nn\\
&\left.\times\f{1}{\sqrt{\k_1^2\k_2^2}}\left(\log\f{\k_1^2}{\k_2^2}
\tan^{-1}\f{|\k_2|}{|\k_1|}+Im~Li_2\left(i\f{|\k_2|}{|\k_1|}
\right)\right)\right]-\f{\pi^2}{3\k^2_>},
\end{align}
in which an azimuthal average has been performed, and the dilogarithmic one
\begin{align}
&\tilde{H}(\k_1,\k_2)+\f{\pi^2}{3\k_>^2}=
\f{2\q\cdot(\k_1+\k_2)}{\q^2(\k_1+\k_2)^2}
\left[\log\f{\k_1^2}{\k_2^2}\log\f{\k_1^2\k_2^2}{(\k_1^2+\k_2^2)^2}+
\right.\nn\\
&\left.+Li_2\left(1-\f{\q^2}{\k_1^2}\right)-
Li_2\left(1-\f{\q^2}{\k_2^2}\right)
+Li_2\left(-\f{\k_2^2}{\k_1^2}\right)-Li_2\left(-\f{\k_1^2}{\k_2^2}\right)
\right]+\nn\\
&+2\left[\int_0^1\f{dt}{(\k_1-t\k_2)^2}\left(\f{\k_2\cdot\q}{\q^2}
-\f{\k_2^2\q\cdot(\k_1+\k_2)}{\q^2(\k_1+\k_2)^2}(1+t)\right)
\log\f{t(1-t)\k_2^2}{\k_1^2(1-t)+\q^2 t}+\right.\nn\\
&\left.+(\k_1\longleftrightarrow-\k_2)\right].
\end{align}

Let us now investigate the physical features emerging from the irreducible NL
kernel, as defined by the sum of Eqs. (6), (12) and (14).
In order to find its eigenvalue, we shall proceed as for the leading term
by applying the kernel to test functions of the form $(\k_2^2)^{\g-1}$,
with $0<\g<1$. Due to its explicit renormalization scale dependence
we expect the outcome to contain factors of $\log(\k_1^2/\mu^2)$.

We combine first the $\ep$-dependent singular terms in Eqs. (12) and (14) and
we notice that the most singular ones ($\sim 1/\ep^2$) cancel out
directly,\footnotemark
\footnotetext{This justifies performing the expansion up to relative
order $\ep^2$, instead of $\ep^3$.}
leaving the total real-emission singular part
\begin{align}
\bks K_{sing}^{(NL)}=&\f{\asb}{\o}\f{\asb}{4}\left[
\f{(\q^2/\mu^2)^\ep}{\q^2}\f{C(\ep)}{\G(1-\ep)}\left[
-\f{11}{6}+\left(\f{67}{18}-\f{\pi^2}{6}\right)\ep
-\left(\f{202}{27}-7\zeta(3)\right)\ep^2+\right.\right.\nn\\
+&\left.\left.\f{11}{3}\f{1}{\ep\q^2\G(1-\ep)}\right]\right].
\end{align}
This kernel has a finite $\ep\ra 0$ limit at fixed $\q^2$, but its
eigenvalues are still singular.

We then compute the eigenvalues of the kernel (17) by using Eq. (4) and the
additional integral
\begin{align}
&\f{C(\ep)}{\G(1-\ep)}\int\f{(\q^2/\mu^2)^\ep}{\q^2}
d[\k_2]\left(\f{\k_2^2}{\k_1^1}\right)^{\g-1}=
\left(\f{\k_1^2}{\mu^2}\right)^{2\ep}\f{1}{\ep^2}
\f{\G^2(1+\ep)\G(\g+\ep)\G(1-\g-2\ep)}{\G^2(1-\ep)\G(\g+3\ep)\G(1-\g)}=
\nn\\
&=\left(\f{\k_1^2}{\mu^2}\right)^{2\ep}\f{1}{\ep^2}
\left[\exp\left(2\ep\chi_0(\g)\right)+\f{1}{2}\ep^2
\left(4\psi^\prime(1-\g)-8\psi^\prime(\g)\right)+O(\ep^3)\right],
\end{align}
and, by combining them with the virtual term (6) we obtain the finite result
\begin{equation}
\f{\asb}{\o}\f{\asb}{4}\left[\chi_0(\g)\left(-\f{11}{3}\log\f{\k_1^2}{\mu^2}+
\f{67}{9}-\f{\pi^2}{3}\right)-\f{11}{6}
\left(\chi_0^2(\g)+\chi_0^\prime(\g)\right)+6\zeta(3)\right].
\end{equation}

Note now that the coefficient of the $\log\mu$ term is
precisely the leading kernel eigenvalue with a beta-function
coefficient. Therefore, it can be interpreted as a running coupling factor,
much as for the $q\bar{q}$ contribution.
We can thus express the total (L+NL) kernel in the form
\begin{align}
K^{(L+NL)}&=\f{\asb(\mu^2)}{\o}\left[\left(
1-b\as(\mu^2)\log\f{\k_1^2}{\mu^2}\right)K_0(\k_1,\k_2)+
\as(\mu^2)K_1(\k_1,\k_2)\right]\nn\\
&\simeq\f{\asb(\k_1^2)}{\o}\left(K_0(\k_1,\k_2)+\as K_1(\k_1,\k_2)\right),
\end{align}
which defines the NL scale-invariant kernel $K_1$.

Factorizing the running coupling at the scale $\k_1^2$ is an asymmetrical
procedure, but is convenient for the discussion of the non-scale-invariant
BFKL equation \cite{13}.
Using a different scale (e. g., $\as(\k_>^2)$) implies changing $K_1$ so as to
leave the total NL kernel invariant.

Finally, a straightforward calculation allows the computation of 
the characteristic function of the remaining finite part of the kernel, 
except for $\tilde{H}$, whose eigenvalue is estimated semi-analytically to be
$\tilde{h}(\g)\simeq\sum_{n=1}^3 a_n[(\g+n)^{-1}+(1-\g+n)^{-1}]$, with
$a_1=.72$, $a_2=.28$, $a_3=.16$.
We thus obtain the gluonic part of the $K_1$ eigenvalue in the form
\begin{align}
&\as\chi_1^{(g)}(\g)=\f{\asb}{4}\left[-\f{11}{6}\left(\chi_0^2(\g)+
\chi_0^\prime(\g)\right)+\left(\f{67}{9}-\f{\pi^2}{3}\right)\chi_0(\g)+\right.
\nn\\
&\left.+\left(6\zeta(3)+\f{\pi^2}{3\g(1-\g)}+\tilde{h}(\g)\right)
-\left(\f{\pi}{\sin\pi\g}\right)^2\f{\cos\pi\g}{3(1-2\g)}
\left(11+\f{\g(1-\g)}{(1+2\g)(3-2\g)}\right)\right],
\end{align}
to be compared with the $N_f$-dependent part obtained previously
\cite{12}
\begin{align}
\!\!\bks\as\chi_1^{(q)}(\g)=\f{N_f\as}{6\pi}\left[
\f{1}{2}\left(\chi_0^2(\g)+\chi_0^\prime(\g)\right)
-\f{5}{3}\chi_0-\f{1}{N_c^2}
\left(\f{\pi}{\sin\pi\g}\right)^2\f{\cos\pi\g}{1-2\g}
\f{1+\f{3}{2}\g(1-\g)}{(1+2\g)(3-2\g)}\right].
\end{align}

Correspondingly, the (azimuthal averaged) $\k$-space kernel can be
rewritten in a more compact form:
\begin{equation}
\as K_1^{(g)}=
\f{\asb}{4}\left[-\f{11}{3\q^2}\log\left.\f{\q^2}{\k_1^2}\right|_R
-H_{coll}(\k_1,\k_2)+\tilde{H}(\k_1,\k_2)+6\pi\zeta(3)\delta^{(2)}(\q)
\right]
\end{equation}
where we have used the notation of Eqs. (15) and (16), and we have introduced
the regularized distributions in $2$-dimensional transverse space
\begin{equation}
\left.f(\k_1,\q)\right|_R=f(\k_1,\q)\Theta(\q^2-\lambda^2)-\delta^2(\q)
\int_{\lambda^2}^{\k_1^2} f(\k_1,\q)d^2\q.
\end{equation}

The parallel expression for the $q\bar{q}$-part, obtained in ref \cite{13},
has the form\footnotemark
\footnotetext{Eq. (3.17) of Ref. \cite{13} contains some misprints,
in particular a $C_f$ factor instead of $N_f$.}
\begin{equation}
\as K_1^{(q)}=\f{N_f\as}{6\pi}\left[\left.\left(\log\f{\q^2}{\k_1^2}
-\f{5}{3}\right)\f{1}{\q^2}\right|_R-\f{1}{N_c^2}H_{ab}(\k_1,\k_2)\right],
\end{equation}
where the abelian contribution is defined by
\begin{equation}
H_{ab}(\k_1,\k_2)=\f{3}{32}\left[\left(\f{1}{\k_2^2}-\f{1}{\k_1^2}
\right)\log\f{\k_1^2}{\k_2^2}+2\left(\f{1}{\k_1^2}+\f{1}{\k_2^2}\right)
\right.+\nn
\end{equation}
\begin{equation}
+\left.\left(22-\f{\k_1^2}{\k_2^2}-\f{\k_2^2}{\k_1^2}\right)
\f{1}{\sqrt{\k_1^2\k_2^2}}
\left(\log\f{\k_1^2}{\k_2^2}\tan^{-1}\f{|\k_2|}{|\k_1|}+
Im~Li_2\left(i\f{|\k_2|}{|\k_1|}\right)\right)\right].
\end{equation}
We notice that in both cases the natural scale of $\as$ appears to be
$\q^2$, rather than $\k_1^2$, and that the "collinear"  and
"abelian" terms have same sign and similar structure, but that the latter is
suppressed by the $1/N_c^2$ colour factor.

Several comments on the results in Eqs. (20)-(26) are in order.
Consider first using them to describe the gluon density in a two-scale
hard process. According to our general arguments \cite{13,14} the
kernel with running coupling is consistent with the renormalization group for
$\k_1^2>\k_2^2\gg\Lambda^2$.

More precisely the Mellin transform $G_\o(Q,Q_0)$ of the BFKL Green's function
is represented by
\begin{equation}
G_\o(Q,Q_0)=\f{1}{\g_+\sqrt{-\chi^\prime(\g_+)}}
\left(\exp\int_{t_0}^{t}\g_+\left(\as(t^\prime)\right)dt^\prime\right)K(\o,t_0)
\end{equation}
in the anomalous dimension regime
\begin{equation}
t=\log\f{Q^2}{\Lambda^2}\gg t_0,~~~~~~b\o t>\chi\left(\f{1}{2}\right),
\end{equation}
where $\g_+\simeq\g_{gg}+\f{C_F}{C_A}\g_{qg}$ is the larger eigenvalue
of the anomalous dimension matrix, defined at NL level by
\begin{equation}
1=\f{\asb(t)}{\o}\left(\chi_0(\g_+)+\as\chi_1(\g_+)\right).
\end{equation}

From the definition (29) both perturbative and resummed expressions of the
NL anomalous dimension follow from the formula
\begin{equation}
\g_+^{NL}(\as,\o)=-\as\f{\chi_1\left(\g_L\left(\f{\asb}{\o}\right)\right)}
{\chi^\prime_0\left(\g_L\left(\f{\asb}{\o}\right)\right)},
\end{equation}
where $\g_L(\asb/\o)=\asb/\o+O(\asb/\o)^4$ is the well-known \cite{2}
leading gluon anomalous dimension.

Therefore, the low order expansion of $\g_+$
\begin{equation}
\g_+=\f{\asb}{\o}+\as\left(A_1+A_2\f{\asb}{\o}+A_3\left(\f{\asb}{\o}\right)^2
+...\right)
\end{equation}
implies the small-$\g$ behaviour of $\chi_1$
\begin{equation}
\chi_1(\g)\simeq \f{A_1}{\g^2}+\f{A_2}{\g}+A_3+O(\g),
\end{equation}
which can be checked on Eqs. (21) and (22) to be consistent with the known
\cite{16} expressions in the DIS scheme
\begin{equation}
\as A_1=-\f{11N_c\as}{12\pi}-\f{N_f\as}{6\pi}\f{1}{N_c^2},~~~~~~~~~~~~~
\as A_2=-\f{N_f\as}{6\pi}\left(\f{5}{3}+\f{13}{6N_c^2}\right).
\end{equation}
For the gluonic part, $H_{coll}$ is responsible for the collinear behaviour,
because $\tilde{H}$ vanishes for either $\k_2^2=0$ or $\k_1^2=0$.

Furthermore, the gluonic eigenvalue provides, through Eq. (30), important
resummation effects which are driven by the negative value
$A_1^{(g)}=-11N_c/12\pi$ at the $\g=1$ double pole in Eq. (21).
This term causes a rapid increase of $\left(-\g^2\chi_1^{(g)}(\g)\right)$
relative to its value at $\g=0$ (Fig. 2(a)), 
which reaches a factor of 2.5 at $\g=1/2$.
Thus, unlike the $q\bar{q}$ part, the gluonic part is expected to be
relevant for scaling violations at HERA.

The second important point concerns the high-energy behaviour expected for the
gluon density. We pointed out in Ref. \cite{14} that in the running $\as$
case, two kinds of critical $\o$-values occur. One is the singularity of
the anomalous dimension expansion occurring in Eq. (29) close to
$\g=1/2$, for which we get an $\as$-dependent value
\begin{align}
\op(\as)&=\asb\left(\chi_0\left(\f{1}{2}\right)+\as\chi_1\left(\f{1}{2}
\right)\right)\nn\\
&=\asb\chi_0\left(\f{1}{2}\right)\left(1-a\asb\right).
\end{align}
The other is the true Pomeron, the $\o$-singularity dominating the
high-energy behaviour beyond the anomalous dimension regime.
While the former admits the rough estimate (34), the latter turns
out to be dependent on the behaviour of $\as$ close to $\k_2^2=\Lambda^2$,
and thus cannot be really predicted.

If we take the formula (34) as a qualitative estimate, we realize that the
NL gluon contributions in Eq. (22) yield a rather large negative shift,
namely $a\simeq 3.4$ with our present knowledge (Fig. 2(b)).
This would mean that the "Pomeron" intercept is substantially decreased,
of the order of $\op\simeq .2$ for $\as=.15$.

Let us emphasize that this indication cannot be taken yet
as a quantitative estimate, because of the scale-dependent contributions to
the kernel that we have neglected and because various cross-checks of the
whole approach are still needed. It means, however, that the NL corrections
go in the direction of bridging the gap with soft physics, by
smoothing out the small-$x$ rise at low values of $Q^2$.

If the above magnitude of NL corrections is confirmed, it raises the
problem of the slow convergence of resummed perturbation theory at small-$x$.
Fortunately some classes of corrections can be roughly understood at all
orders, because they correspond to physical phenomena we already know about.

One class of corrections is due to the collinear behaviour of large-$x$
contributions, which in the cluster expansion approach occur in higher
order clusters, and give rise to multiple poles of $\chi(\g)$ at
$\g=0$ and $\g=1$. Resumming these poles is mandatory \cite{13} to
understand the lower eigenvalue of the anomalous dimension, and in general the
behaviour of $\chi(\g)$ close to $\g=0$.
To see the point, we have plotted in Fig. 4 the function
\begin{equation}
2\psi(1)-\psi(\g-A_1\as)-\psi(1-\g-A_1\as)+\as\left(\chi_1(\g)-A_1
\left(\f{\pi}{\sin\pi\g}\right)^2\right)
\end{equation}
which coincides up to NL level with $\chi_0+\as\chi_1$, but
differs at higher orders, by a resummation of the collinear behaviour.

It is apparent that the effect of resummation is to displace the
$\g=0$ singularity to $\g=A_1\as<0$, as expected, and to reduce the
Pomeron shift, by about $15\%$ for $\as=.15$.

Another class of corrections was noticed long ago by one of us \cite{17}
to be due to coherence effects in the soft gluon emission region.
The ensuing structure function equation with angular ordering,
further investigated by Catani, Fiorani and Marchesini \cite{18}
(the CCFM equation), is the basis for the treatment of such effects to
all orders.

From the results which are already available \cite{19}, it appears that
coherence plays a role starting from the constant $A_3$ in Eq. (32), and thus
from the three loop level. It constitutes, therefore,a rather delicate check
of the whole approach, because the scale-dependent terms,
neglected here, are also expected to contribute.

On the whole, we think that a full understanding of the next-to-leading
kernel will put several phenomenological issues on quantitative grounds and
will help to bridge the gap with large-$x$ properties, low $Q^2$ physics,
and diffractive phenomena.

\begin{center}
{\bf Acknowledgements}
\end{center}

One of us (M. C.) is grateful to Yuri Dokshitzer, Victor Fadin, Al Mueller,
Giuseppe Marchesini and Gavin Salam for a quite stimulating discussion.
We thank Jochen Bartels, Alan Martin and Zoltan Kunszt for interesting
discussions and suggestions.
This work is supported in part by E. C. contract \# CHRX-CT96-0357 and
by MURST (Italy).

\newpage
{\large \bf Figure captions:}\\ 

{\bf Figure 1:} (a-c) Leading and (d-f) next-to-leading contributions to the 
high energy cluster expansion.
The loop number of radiative corrections is specified, and clusters with
external particles are omitted. Intermediate quarks and gluons are understood 
in (f).

{\bf Figure 2:} Plot of (a) the gluonic contribution to $\g^2\chi_1(\g)$ and 
(b) the $\as$-dependent "Pomeron".

{\bf Figure 3:} Plot of the resummed characteristic function in Eq. (35), 
with the symmetrical choice $\chi_1^>(\g)$, for which $\as(\k^2_>)$ is 
factorized. The corresponding Pomeron shift should be decreased by
$\Delta\op=-\f{b}{2}\as\asb\pi^2$ to compare with the result in Eq. (34).

\newpage

\begin{figure}
\vspace*{-5 cm}\hspace*{-1 cm}
\centerline{\includegraphics[width=16 cm]{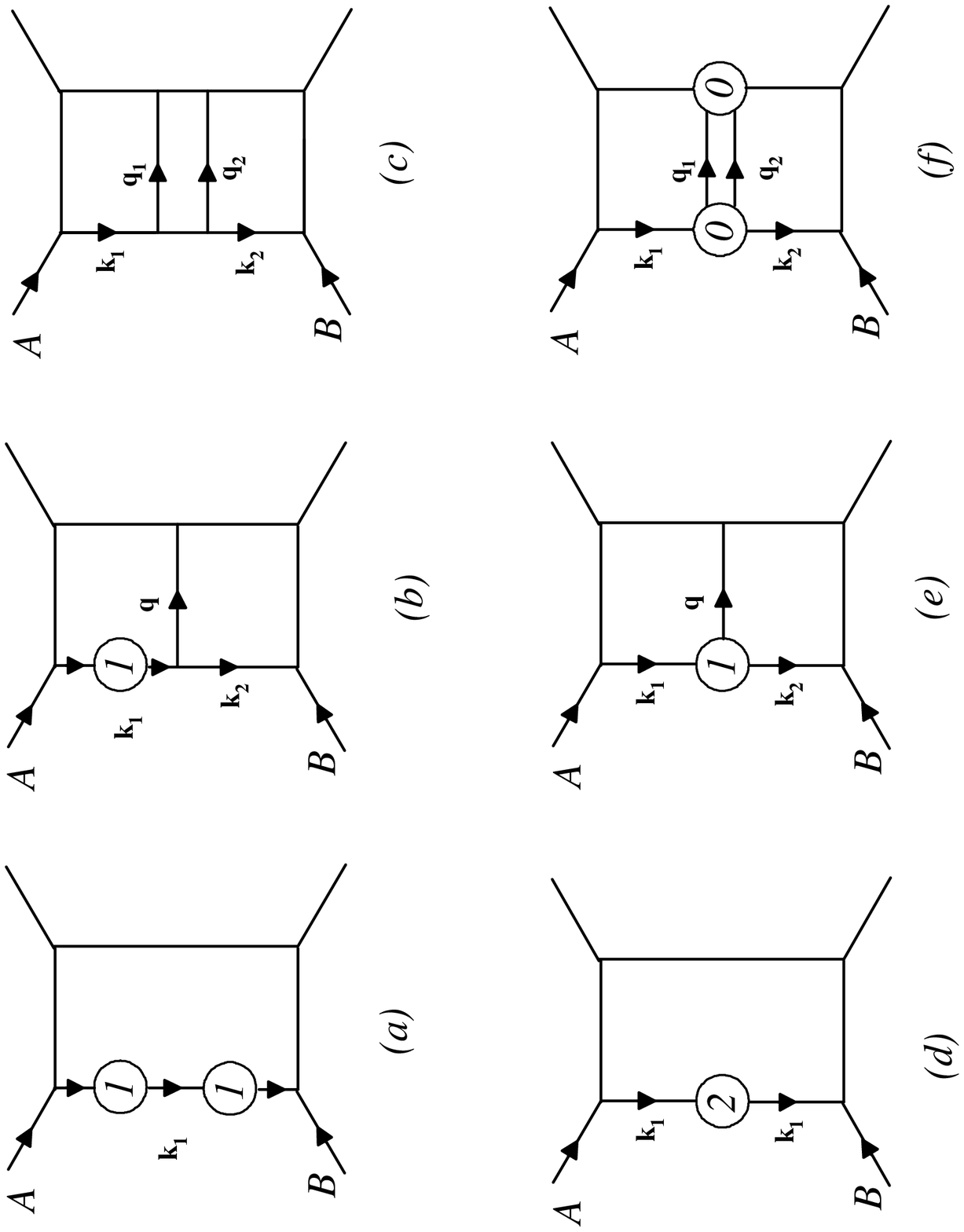}}
\end{figure}

\newpage
\begin{figure}
\hspace*{5 cm}\vspace*{12 cm}
\centerline{\includegraphics[angle=90,width=26 cm]{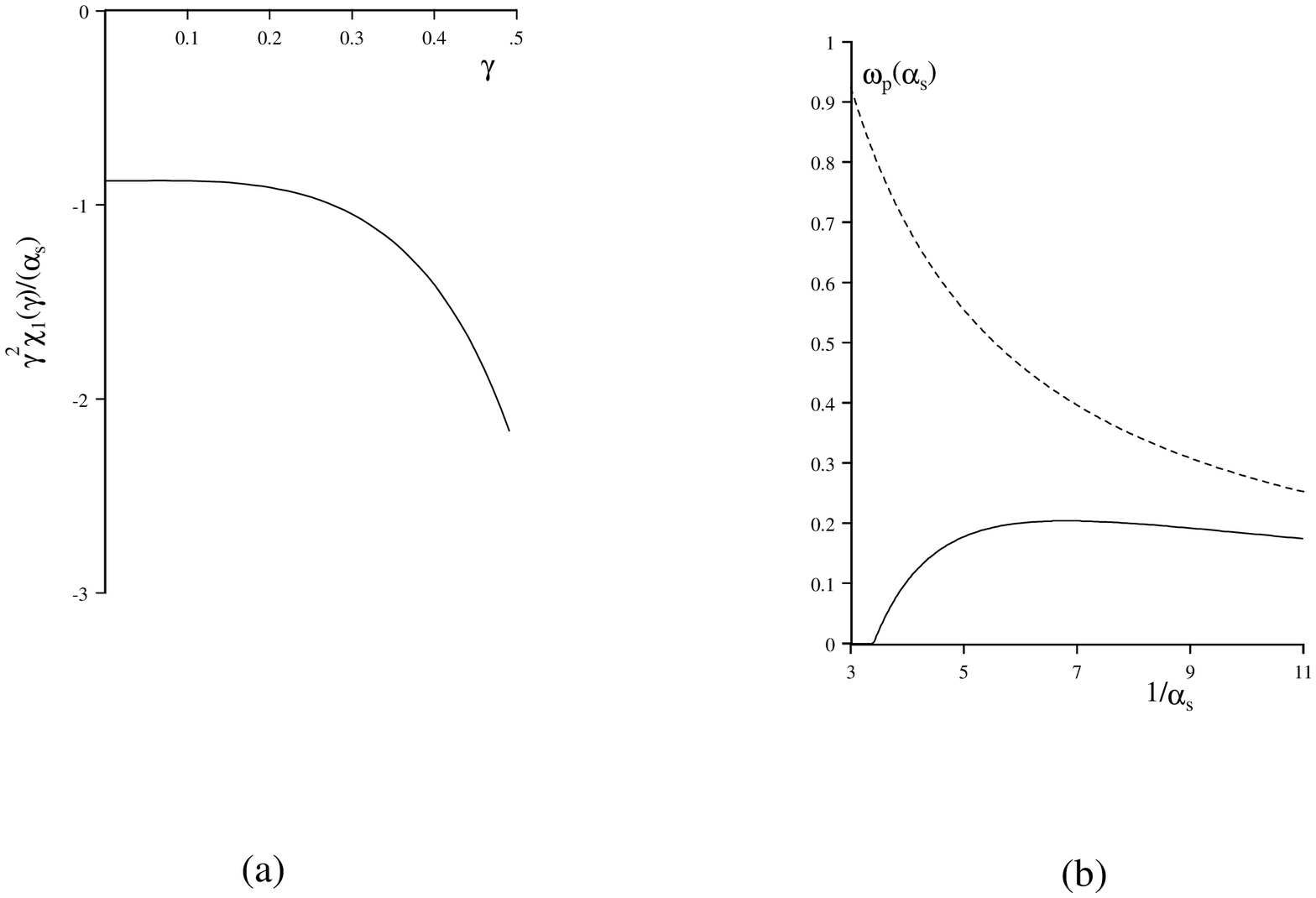}}
\vspace*{-8 cm}
\end{figure}

\newpage
                                          
\begin{figure}
\vspace*{-5 cm}
\centerline{\includegraphics{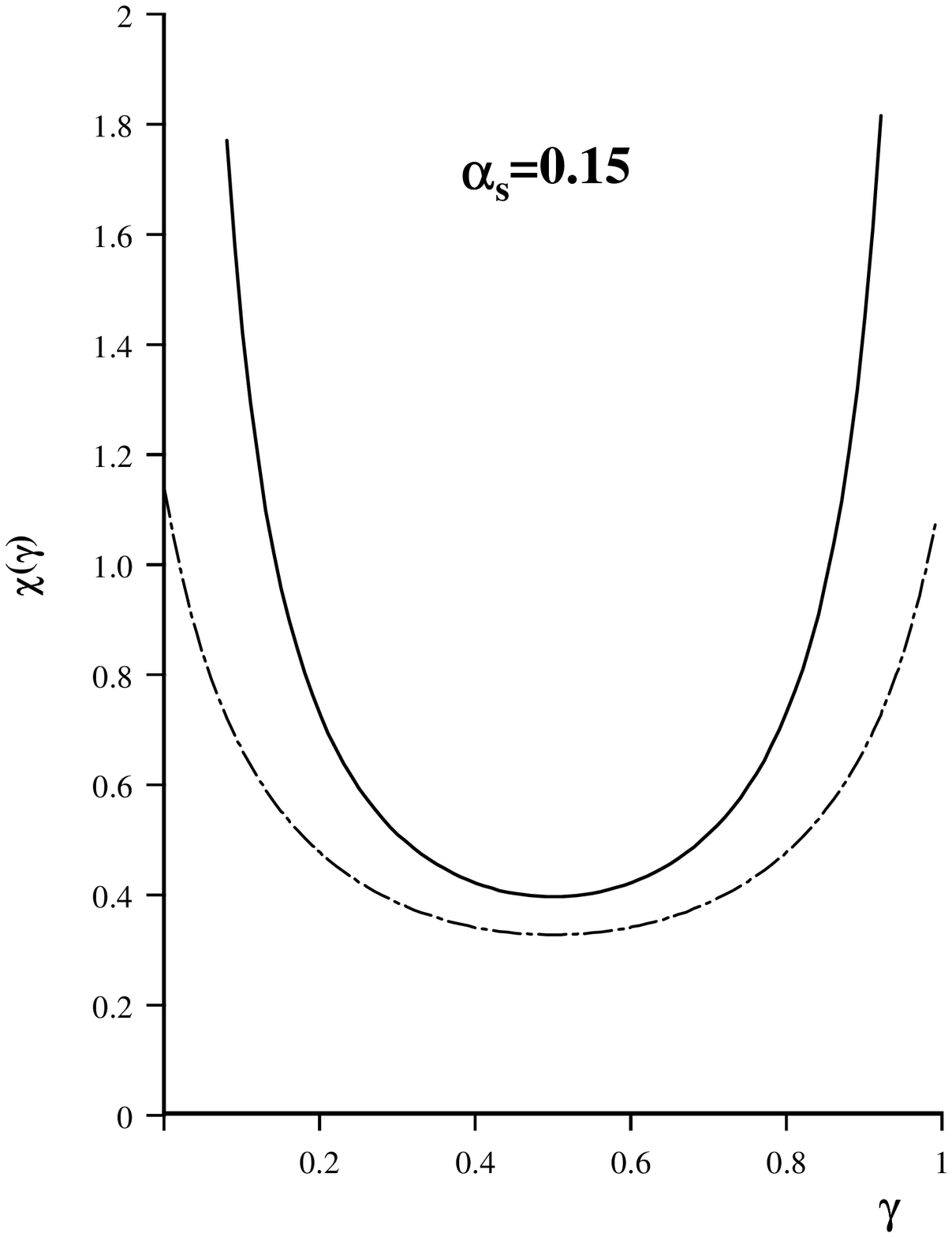}}
\vspace*{-3cm}
\end{figure}


\begin{thebibliography}{99}
\bibitem{1} S. Aid et al. H1 Collaboration, Nucl. Phys. B 470 (1996) 3;\\
The ZEUS Collaboration, Z. Phys. C 69 (1996) 607.
\bibitem{2} L.N. Lipatov, Sov. J. Nucl. Phys. 23 (1976) 338;\\ 
E.A. Kuraev, L. N. Lipatov and V. S. Fadin Sov. Phys. JETP 45 (1977) 199;\\ 
Ya. Balitskii and L. N. Lipatov, Sov. J. Nucl. Phys. 28 (1978) 822.
\bibitem{3} V. S. Fadin and L. N. Lipatov, Yad. Fiz. 50 (1989) 1141.
\bibitem{4} S. Catani, M. Ciafaloni and F. Hautmann, Phys. Lett. B 242 (1990)
97; Nucl. Phys. B 366 (1991) 135.
\bibitem{5} V. S. Fadin and L. N. Lipatov, Nucl. Phys. B 406 (1993) 259.
\bibitem{6} V. S. Fadin, R. Fiore and A. Quartarolo, Phys. Rev. D 50 (1994) 
2265; Phys. Rev. D 50 (1994) 5893.
\bibitem{7} V. S. Fadin, R. Fiore and M. I. Kotsky, Phys. Lett. B 389 (1996)
737.
\bibitem{8} V. S. Fadin, R. Fiore and M. I. Kotsky, Phys. Lett. B 359 (1995)
181 and Phys. Lett. B 387 (1996) 593.
\bibitem{9} V. S. Fadin and L. N. Lipatov, Nucl. Phys. B 477 (1996) 767;
\bibitem{10} V. del Duca, Phys. Rev. D 54 (1996) 989; Phys. Rev. D 54
(1996) 4474. 
\bibitem{11} V. S. Fadin, L. N. Lipatov and M. I. Kotsky,
preprint hep-ph/9704267.
\bibitem{12} G. Camici and M. Ciafaloni, Phys. Lett. B 386 (1996) 341.
\bibitem{13} G. Camici and M. Ciafaloni, Nucl. Phys. B 496 (1997) 305.
\bibitem{14} G. Camici and M. Ciafaloni, Phys. Lett. B 395 (1997) 118.
\bibitem{15} V. S. Fadin, talk given at the international conference
"Deep Inelastic Scattering 96", Roma, April 1996.
\bibitem{16} G. Curci, W. Furmanski and R. Petronzio, Nucl. Phys. B 175 (1980)
27; E. B. Zijlstra and W. L. van Neerven, Nucl. Phys. B383 (1992) 525.
\bibitem{17} M. Ciafaloni, Nucl. Phys. B 296 (1988) 49, in particular
Appendix C.
\bibitem{18} S. Catani, F. Fiorani and G. Marchesini, Phys. Lett. B 234 (1990)
339, Nucl. Phys. B 336 (1990) 18.
\bibitem{19} G. Bottazzi, G. Marchesini, G. P. Salam and M. Scorletti,
preprint hep-ph/9702418.

\end{thebibliography}
\end{document}